\documentclass{article}
\begin{document}
\title{New derivation of the cluster cumulant formula}
\author{D K Sunko\thanks{email: dks@phy.hr}\\
Department of Physics,\\Faculty of Science,\\
University of Zagreb,\\
Bijeni\v cka cesta 32,\\
HR-10000 Zagreb, Croatia.}
\date{}

\maketitle
\begin{abstract}
The cluster cumulant formula of Kubo is derived by appealing only to
elementary properties of subsets and binomial coefficients. It is shown to be
a binomial transform of the grand potential. Extensivity is proven without
introducing cumulants. A combinatorial inversion is used to reformulate the
expansion in the activity to one in occupation probabilities, which
explicitly control the convergence. The classical virial expansion is
recovered to third order as an example.
\end{abstract}


\section{Introduction}

The classical (Ursell-Yvon-Mayer) virial expansion is the traditional
introduction to real-space particle correlations in basic statistical physics
textbooks, both old~\cite{Huang63} and new~\cite{Balian91}. By contrast,
cumulant methods, of which the virial expansion is a special case, are left
to more specialized texts~\cite{Fulde93}. A possible disadvantage of this
approach is that physical issues are initially confused with purely formal
ones, stemming from the specific expression for the grand potential
$\Omega=-kT\Psi$ as a logarithm of a sum:
\begin{equation}
\Psi=\ln\sum_Ne^{\beta\mu N}\mathrm{tr}_N\exp(-\beta H).
\label{def}
\end{equation}
The most prominent formal problem is extensivity: it is important to
demonstrate that the `factorization property' of matrix elements leads to a
$\Psi$ proportional to the number of single-particle states.

The formal and physical side of the problem were neatly separated by Kubo
\cite{Kubo62}. This article is built around a very compact formal derivation
of one of his main formulas, appealing only to elementary algebra. Its
principal feature is that the usual ordering, by which the cluster cumulant
formula is a rearrangement of the cumulant expansion, is reversed, and the
cluster formula appears as the basic one. At the level considered here,
operative formulas can be developed without invoking cumulants at all,
physics being determined by the choice of expansion parameter. Given the
importance of cumulants in general, this is not necessarily a pedagogical
advantage. It is nevertheless hoped that the present work gives a reasonable
introduction to the issues involved in cumulant expansions, at a level still
attainable in a general (graduate) course of statistical physics. To this
end, the virial series is rederived as an example, not because this is the
quickest way to do it (it obviously is not), but in order to provide a list,
in some logical order, of the steps involved in reducing the fully general
quantum formula to an operative classical limit. Once such a framework is
established, each of these steps can be made the starting point of a more
specialized development.

Being concerned with interpretation, rather than calculation, this article
cannot do justice to the numerous, sometimes very refined, applications of
the cumulant approach, in its broadest sense, which have developed over a
period of more than half a century. Outside the original context of a
classical gas~\cite{Royer80}, perhaps the most detailed work was done on spin
models, under the heading of high-, and low-, temperature
expansions~\cite{Domb74}. Another long-standing field of applications are
polymers, including fundamental issues, such as the excluded-volume problem
far from the gelation point~\cite{Barrett94}. Strong electron
correlations~\cite{Becker90,Sunko96,Polatsek97} have also been described by
cumulant expansions.  The common denominator of all these applications is the
need to treat correlations in real space, when neither scaling nor
periodicity can be used to simplify the problem. An interesting variant of
the cumulant method is when a finite system is treated in some `inverse'
space, which diagonalizes part of the Hamiltonian; this is a classic
microscopic approach in nuclear physics~\cite{Blaizot86}.

Concerning the presentation, I am aware that combinatorial manipulations are
not part of the usual education of a physicist. If the binomial transform
were as commonly known as the Fourier transform, the main result in the next
section would be a `one-liner'. I have tried to spell everything out, using
familiar set-theoretic notation. A word of advice to the diligent reader:
follow the formulas on a small example, with two, or at most three,
single-particle states. If it suddenly appears easy, that is because it is.

\section{The Kubo formula for fermions}

\subsection{The activity expansion}

The diagonal matrix element in $N$-particle space is denoted
\begin{equation}
U_{i(N)}\equiv \left<i_1,\ldots,i_N\right|e^{-\beta H}
\left|i_1,\ldots,i_N\right>,
\end{equation}
where a set of occupied single-particle states (configuration) is denoted
by  $i(N)\equiv\{ i_1,\ldots,i_N\}$. It is assumed that the total number of
single-particle states is $L$, and they comprise a set, to be denoted
$\mathcal{L}$. In this set-theoretic notation, the Massieu potential
(\ref{def}) reads
\begin{equation}
\Psi=\ln\sum_{\emptyset\subseteq i(N)\subseteq\mathcal{L}(L)}
e^{\beta\mu N}U_{i(N)}.
\end{equation}
The sum is over all subsets of of the set $\mathcal{L}$ of single-particle
states, which may of course take at most $L$ fermions. It will be assumed
throughout that the vacuum expectation $U_\emptyset=1$. All the derivations in
this article will refer to fermions. Only the final expressions for bosons are
given in the Appendix.

The motivation for the next step is that one would like a sum outside the
logarithm, not inside it. So define an intermediate expression,
\begin{equation}
\Psi_M\equiv\sum_{j(M)\subseteq\mathcal{L}}
\ln\sum_{\emptyset\subseteq i(N)\subseteq j(M)}e^{\beta\mu N}U_{i(N)}.
\label{psiM}
\end{equation}
Here the sum inside is limited to some subset of $\mathcal{L}$, with exactly
$M$ states, and the sum outside is over all possible choices of such a
subset. Notice that $\Psi=\Psi_L$, since the sum outside the logarithm then
reduces to a single term, $j(L)=\mathcal{L}$.

Now introduce the pair of inverse relations
\begin{eqnarray}
\Psi_M&=&\sum_{m=0}^M{L-m\choose M-m} \Psi(m),\label{prva}\\
\Psi(m)&=&\sum_{M=0}^m(-1)^{m-M}{L-M\choose m-M} \Psi_M.\label{druga}
\end{eqnarray}
If one of these is inserted into the other, the result is an identity, so
they are valid independently of what the $\Psi_M$ may be. Taking (\ref{prva})
at $M=L$, one finds
\begin{equation}
\fbox{$\displaystyle
\Psi=\sum_{m\ge 1}^{\vphantom{m\ge 1}}\Psi(m),
$}
\label{razvoj}
\end{equation}
which is the Kubo cluster cumulant expansion, when $\Psi(m)$ is given in
terms of the $\Psi_M$ in equation (\ref{psiM}). (Note that
$\Psi(0)=\Psi_0=\ln U_\emptyset=0$.)

To see the structure of (\ref{razvoj}), look at the first two terms, reverting
to standard notation for a moment:
\begin{eqnarray}
\Psi(1)&=&\sum_{i=1}^L\ln\left(1+e^{\beta\mu}\left<i\right|e^{-\beta H}\left|i\right>
\right),\\
\Psi(2)&=&\sum_{1\le i< j\le L}\ln
\left(1+e^{\beta\mu}\left<i\right|e^{-\beta H}\left|i\right>+
e^{\beta\mu}\left<j\right|e^{-\beta H}\left|j\right>+
e^{2\beta\mu}\left<i,j\right|e^{-\beta H}\left|i,j\right>
\right)\nonumber\\
&&-(L-1)\sum_{i=1}^L
\ln\left(1+e^{\beta\mu}\left<i\right|e^{-\beta H}\left|i\right>
\right)\nonumber\\
&=&\sum_{1\le i< j\le L}\left[
\ln\left(1+e^{\beta\mu}\left<i\right|e^{-\beta H}\left|i\right>+
e^{\beta\mu}\left<j\right|e^{-\beta H}\left|j\right>\right.+
e^{2\beta\mu}\left<i,j\right|e^{-\beta H}\left|i,j\right>
\right)\nonumber\\
&&\left.
-\ln\left(1+e^{\beta\mu}\left<i\right|e^{-\beta H}\left|i\right>\right)
-\ln\left(1+e^{\beta\mu}\left<j\right|e^{-\beta H}\left|j\right>\right)
\right]\nonumber\\
&=&\sum_{1\le i< j\le L}\ln\left(
1+e^{2\beta\mu}\frac{
\left<i,j\right|e^{-\beta H}\left|i,j\right>-
\left<i|e^{-\beta H}|i\right>\left<j\right|e^{-\beta H}\left|j\right>}
{\left(1+e^{\beta\mu}\left<i\right|e^{-\beta H}\left|i\right>\right)
\left(1+e^{\beta\mu}\left<j\right|e^{-\beta H}\left|j\right>\right)}\right).
\end{eqnarray}
This should be compared with equation 6.22 of reference~\cite{Kubo62}. The
lowest power of the activity appearing in $\Psi(m)$ is $e^{\beta\mu m}$, so I
shall refer to this as the `activity expansion'. If the basis
$\left|i\right>$ diagonalizes the Hamiltonian, $\Psi(1)$ gives the exact
solution of the trivial problem, and the other $\Psi(m)$ are zero.

Of the three forms in which $\Psi(2)$ is given, the first is the defining
form (\ref{druga}). The second is Kubo's original form, where all terms
appear under a single sum. The original form reads (equation 4.13 of
reference~\cite{Kubo62})
\begin{equation}
\Psi(m)=\sum_{i(m)\subseteq\mathcal{L}}\psi[i(m)],
\label{psiorig}
\end{equation}
where
\begin{equation}
\psi[i(m)]\equiv\sum_{\emptyset\subseteq j(M)\subseteq i(m)}(-1)^{m-M}
\ln\left[
\sum_{\emptyset\subseteq k(l)\subseteq j(M)}e^{l\beta\mu}U_{k(l)}\right].
\label{psi}
\end{equation}
Finally, the third is a compact form, where all logarithms have been
collapsed to a single one. The trouble with it, although it looks
prettiest for small $m$, is a large (hyperexponential) explosion with $m$ in
the number of terms in the fraction multiplying $e^{\beta\mu m}$, as the
fraction arises from multiplication of the polynomials in $e^{\beta\mu}$
which appear under the logarithms in the original form.

It is easy to prove that the defining and original forms are equivalent. One
starts from equation (\ref{psiorig}) and simply interchanges the order of
summation:
\begin{eqnarray}
\Psi(m)&=&\sum_{i(m)\subseteq\mathcal{L}}
\sum_{\emptyset\subseteq j(M)\subseteq i(m)}(-1)^{m-M}\ln[\ldots j(M)\ldots]
\nonumber\\
&=&\sum_{\emptyset\subseteq j(M)\subseteq\mathcal{L}}
(-1)^{m-M}\ln[\ldots j(M)\ldots]
\sum_{j(M)\subseteq i(m)\subseteq\mathcal{L}} 1\nonumber\\
&=&\sum_{\emptyset\subseteq j(M)\subseteq\mathcal{L}}
(-1)^{m-M}\ln[\ldots j(M)\ldots]{L-M\choose m-M}\nonumber\\
&=&\sum_{M\ge 0}(-1)^{m-M}{L-M\choose m-M}\Psi_M,\nonumber
\end{eqnarray}
where, in the last step, all $M$-particle configurations have been grouped
together. From this point of view, the present article rests on the
observation that the last line above is formally invertible.

\subsection{Extensivity}

Extensivity itself cannot, of course, be proven without some reference to the
interactions involved. What will be proven here is more properly called
size-consistency: if the matrix element $U_{i(m)}$ in a configuration
$i(m)$ can be expressed as a product of lower matrix elements, then the
contribution of that configuration to the grand potential is zero.

We shall need an elementary property of sets. Namely, if a set $i(m)$ is
written as the union of two non-overlapping, non-empty subsets,
$i(m)=i_1(m_1)\cup i_2(m_2)$, then all the subsets of $i(m)$ may be obtained
by writing down all subsets of $i_1(m_1)$ and $i_2(m_2)$, and combining them
in all possible ways. In particular, if there is a sum over subsets of
$i(m)$, it can be written as two sums:
\begin{equation}
\sum_{\emptyset\subseteq j(n)\subseteq i(m)}
\left[\ldots j(n)\ldots\right]=
\sum_{\emptyset\subseteq j_1(n_1)\subseteq i_1(m_1)}
\sum_{\emptyset\subseteq j_2(n_2)\subseteq i_2(m_2)}
\left[\ldots j_1(n_1)\cup j_2(n_2)\ldots\right].
\end{equation}
This will be referred to below as the `subset property'.

The proposition is as follows: let $i(m)$ be the union of two
non-overlapping, non-empty subsets, $i(m)=i_1(m_1)\cup i_2(m_2)$. Let
\begin{equation}
U_{k_1(l_1)\cup k_2(l_2)}=U_{k_1(l_1)}U_{k_2(l_2)}
\end{equation}
whenever $k_1(l_1)\subseteq i_1(m_1)$ and $k_2(l_2)\subseteq i_2(m_2)$. Then
$\psi[i(m)]=0$, where $\psi[i(m)]$ is the contribution of the set
$i(m)$ to $\Psi(m)$ in (\ref{psiorig}). 

For the proof, first observe that under these assumptions, the sum under the
logarithm in (\ref{psi}) factorizes:
\begin{equation}
\sum_{\emptyset\subseteq k(l)\subseteq j(n)}U_{k(l)}=
\sum_{\emptyset\subseteq k_1(l_1)\subseteq j_1(n_1)}U_{k_1(l_1)}
\sum_{\emptyset\subseteq k_2(l_2)\subseteq j_2(n_2)}U_{k_2(l_2)}.
\label{fact}
\end{equation}
This is trivial: the sum is rewritten by the subset property, and the matrix
element factorizes by assumption. It follows that the logarithm of
(\ref{fact}) is the sum of two logarithms, one a function of $j_1(n_1)$
alone, the other of $j_2(n_2)$. The contribution of the first logarithm to
(\ref{psi}) reads
\begin{eqnarray}
\lefteqn{\sum_{\emptyset\subseteq j(n)\subseteq i(m)}(-1)^{m-n}
\ln\left[\ldots j_1(n_1)\ldots\right]}&&\\
&=&\sum_{\emptyset\subseteq j_1(n_1)\subseteq i_1(m_1)}(-1)^{m_1-n_1}
\ln\left[\ldots j_1(n_1)\ldots\right]
\sum_{\emptyset\subseteq j_2(n_2)\subseteq i_2(m_2)}(-1)^{m_2-n_2},\nonumber
\end{eqnarray}
because the sum here can also be written by the subset property.
Since the number of subsets of $i(m_2)$ with fixed number $n_2$ is given by a
binomial coefficient,
\begin{equation}
\sum_{\emptyset\subseteq j_2(n_2)\subseteq i_2(m_2)}(-1)^{m_2-n_2}=
\sum_{n_2=0}^{m_2}{m_2\choose n_2}(-1)^{m_2-n_2}=(1-1)^{m_2},
\end{equation}
this is zero for $m_2\ge 1$; similarly the contribution of the second
logarithm is zero for $m_1\ge
1$, so the proposition is proved.

\section{The probability expansion}

In this section, the combinatorial approach is pushed a step further, to
rewrite Kubo's formula (\ref{razvoj}) in a particularly transparent way. (The
derivations are written somewhat more tersely than in the other sections.)
Probabilities will replace the activity as the expansion parameters,
analogously to passing from activity to concentration in the classical case.
It should be emphasized that this only affects the form of the $\psi[i(m)]$
in (\ref{psiorig}), their value remaining the same, term for term, for all
$i(m)$ with $m\ge 2$. The simplest example of the transformation is in the
two ways one may write $\Psi(1)$, 
\begin{equation}
\Psi(1)=\sum_i\ln\left(1+e^{\beta\mu}U_i\right)=
-\sum_i\ln\left(1-p_i\right), \label{nonint} \end{equation} where the
`occupation probability' \begin{equation} p_i={e^{\beta\mu}U_i\over
1+e^{\beta\mu}U_i} \label{pi} \end{equation} would be just the Fermi function
in the quantum non-interacting case.

\subsection{The general transformation}

Here a whole class of ways to expand $\Psi$ will be shown to be equal to the
activity expansion (\ref{razvoj}), term for term. In other words, they are
merely different ways to rearrange the contributions under the logarithms in
(\ref{psi}). A special choice then gives the expansion in the probabilities
(\ref{pi}), alluded to above.

Take an arbitrary set of $L$ variables $\varepsilon_i$, $i=1,\ldots,L$. Define
the quantities $W$ by the pair of inverse relations
\begin{eqnarray}
U_{i(m)}&=&e^{-\beta\sum_{i\in i(m)}\varepsilon_i}
\sum_{\emptyset\subseteq j(n)\subseteq i(m)} W_{j(n)},\\
W_{j(n)}&=&\sum_{\emptyset\subseteq i(m)\subseteq j(n)}(-1)^{m-n}
e^{+\beta\sum_{i\in i(m)}\varepsilon_i}U_{i(m)}.
\end{eqnarray}
One could just write $x_i$ instead of $e^{-\beta\varepsilon_i}$, but
the notation is meant to be suggestive. The grand partition function may
now be written
\begin{equation}
\sum_{\emptyset\subseteq i(N)\subseteq\mathcal{L}(L)}
e^{\beta\mu N}U_{i(N)}=
\prod_{i=1}^L\left(1+e^{\beta(\mu-\varepsilon_i)}\right)
\sum_{\emptyset\subseteq j(n)\subseteq\mathcal{L}(L)}
f_{j_1}\ldots f_{j_n}W_{j(n)}.
\label{part}
\end{equation}
Here the $f_i$'s are just the Fermi functions corresponding to the
$\varepsilon_i$. Obviously, they are the new variables of the partition
function, replacing the activity. Now define, by analogy with equation
(\ref{psiM}),
\begin{equation}
\widetilde{\Psi}_M\equiv\sum_{i(M)\subseteq\mathcal{L}}
\ln\sum_{\emptyset\subseteq j(n)\subseteq i(M)}f_{j_1}\ldots f_{j_n}W_{j(n)},
\end{equation}
and the main statement of this section is as follows:
\begin{equation}
\widetilde{\Psi}(m)=\Psi(m),\;m\ge 2,
\end{equation}
where $\widetilde{\Psi}(m)$ is the binomial transform (\ref{druga}) of
$\widetilde{\Psi}_M$. In other words, the activity expansion is unaffected by the
transformation, except in the first term; it is easy to show that
\begin{equation}
\sum_i\ln\left(1+e^{\beta(\mu-\varepsilon_i)}\right)
+\widetilde{\Psi}(1)=\Psi(1),
\end{equation}
with the $\varepsilon_i$'s cancelling exactly. Of course, since the partition
function [left-hand side of equation (\ref{part})] does not depend on them,
they all must cancel in the end; but the statement here is that they do so
term by term in the expansion, when $m\ge 2$.

To prove this, express $W$ in $\widetilde{\Psi}_M$ back in terms of the
$U$'s:
\begin{equation}
\sum_{\emptyset\subseteq j(n)\subseteq i(M)}f_{j_1}\ldots
f_{j_n}W_{j(n)}
=\prod_{i\in i(M)}\left(1+e^{\beta(\mu-\varepsilon_i)}\right)^{-1}
\sum_{\emptyset\subseteq k(N)\subseteq i(M)}e^{\beta\mu N}U_{k(N)}.
\end{equation}
Now observe that the sum is the same one as appears in the definition
(\ref{psiM}) of $\Psi_M$, by which $\Psi(m)$ is defined. The difference
$\Psi(m)-\widetilde{\Psi}(m)$ is thus due to the product in front, and reads
explicitly, by equation (\ref{druga}),
\begin{equation}
\sum_M(-1)^{m-M}{L-M\choose m-M}\sum_{i(M)\subseteq\mathcal{L}}
\sum_{i\in i(M)}\ln\left(1+e^{\beta(\mu-\varepsilon_i)}\right),
\end{equation}
which is, in fact, zero for $m\ge 2$. Namely, in the sum over configurations
(subsets of $\mathcal{L}$), each \emph{given} single-particle state will
appear ${L-1\choose M-1}$ times, so upon exchanging the last two sums, one
gets
\begin{equation}
\left[\sum_{i=1}^L\ln\left(1+e^{\beta(\mu-\varepsilon_i)}\right)
\right]\sum_M(-1)^{m-M}{L-M\choose m-M}{L-1\choose M-1},
\end{equation}
and after rearranging the binomial coefficients, the last sum is equal to
\begin{equation}
{L-1\choose m-1}\sum_M(-1)^{m-M}{m-1\choose M-1}\sim (1-1)^{m-1},
\end{equation}
so the statement is proven.

\subsection{The probability expansion}

By choosing
\begin{equation}
e^{-\beta\epsilon_i}=\left<i\right|e^{-\beta H}\left|i\right>=U_i,
\label{onep}
\end{equation}
one finds $f_i=p_i$ [equation (\ref{pi})], and $W_i=0$. This is the useful
case, so let us denote the $W$'s for this special choice by the letter $S$,
for `subtracted':
\begin{equation}
S_{i(m)}=\sum_{\emptyset\subseteq j(n)\subseteq i(m)}(-1)^{n-m}
\widetilde{U}_{j(n)},
\label{sdef}
\end{equation}
where
\begin{equation}
\widetilde{U}_{j(n)}={U_{j(n)}\over U_{j_1}\cdots U_{j_n}},\;
\widetilde{U}_{\emptyset}=1.
\label{normu}
\end{equation}
In terms of these, the $\psi[i(m)]$ in (\ref{psi}) read, for $m\ge 2$,
\begin{equation}
\psi[i(m)]=\sum_{\emptyset\subseteq j(n)\subseteq i(m)}(-1)^{m-n}
\ln\left[
\sum_{\emptyset\subseteq k(l)\subseteq j(n)}p_{k_1}\ldots p_{k_l}S_{k(l)}
\right],
\label{prob}
\end{equation}
noting that $S_\emptyset=1$. This is the `probability expansion'. For
example,
\begin{equation}
S_{ij}=\frac{U_{ij}}{U_iU_j}-1,\;\Psi(2)=\sum_{1\leq i<j\leq L}
\ln(1+p_ip_jS_{ij}),
\end{equation}
and even $\Psi(3)$ is short, in the compact form:
\begin{eqnarray}
\Psi(3)&=&\sum_{1\leq i<j<k\leq L}\ln\left[1+p_ip_jp_k
\vphantom{\frac{
S_{ijk}-p_iS_{ij}S_{ik}-p_jS_{ij}S_{jk}-p_kS_{ik}S_{jk}-p_ip_jp_k
S_{ij}S_{ik}S_{jk}}{
(1+p_ip_jS_{ij})(1+p_ip_kS_{ik})(1+p_jp_kS_{jk})}}\right.
\nonumber\\
&&\left.\times\frac{
S_{ijk}-p_iS_{ij}S_{ik}-p_jS_{ij}S_{jk}-p_kS_{ik}S_{jk}-p_ip_jp_k
S_{ij}S_{ik}S_{jk}}{
(1+p_ip_jS_{ij})(1+p_ip_kS_{ik})(1+p_jp_kS_{jk})}\right]\!\!.
\label{psi3}
\end{eqnarray}
These are higher-order corrections to the non-interacting (`undergraduate')
formula (\ref{nonint}). It is obvious how the probabilities control the
convergence.

It should be noted that the `compact' form has not become really compact, but
rather that the hyperexponential explosion takes off a little later (because
$S_i=0$, so the polynomials being multiplied are shorter). For instance, the
numerator in the compact form of $\Psi(4)$ has $15\,629$ terms in the
activity expansion, and `only' $505$ terms in the probability expansion,
still far less practical than the $15$ distinct terms, $65$ additions and
$16$ logarithms in the original form.

\section{Example}

In this section, the classical virial expansion will be obtained term by
term, from the formulas developed so far. While such a derivation is nothing
new in itself, the purpose is to comment on it from the present
`combinatorial' point of view, and give a familiar interpretation of the
subtracted matrix elements, formally introduced in the previous section.

\subsection{The classical limit}

Begin with a Hamiltonian of the form $H=K+V$, where $K$ is the usual kinetic
energy, and $V$ a sum of two-body interactions depending on mutual distance.
The first choice to be made in the formal expansion is, which basis to use
for the single-particle states. If the momentum basis, which diagonalizes
$K$, is used as a starting point, the term $\Psi(1)$ will be the exact
solution of the non-interacting problem, valid down to zero temperature, and
the corrections will correspond to an expansion in quasiparticle occupation
probabilities, after a canonical transformation to particles and holes. (If
this transformation is not made, the occupation probabilities of levels below
the Fermi energy tend to unity, leading to convergence problems.) Such an
example, inspired by nuclear physics~\cite{Blaizot86}, is beyond the scope of
the present work.

If, on the other hand, the position basis is chosen, one is led straight to
the Mayer expansion. This is particularly easy to see on a lattice. The
one-particle matrix element is then just the normalized one-particle
partition function, independently of position:
\begin{equation}
U_i=\left<\mbox{\boldmath $r$}_i\left|e^{-\beta(K+V)}\right|
\mbox{\boldmath $r$}_i\right>={1\over L}\sum_{\mbox{\boldmath $k$}}
\exp\left[-\beta\varepsilon(\mbox{\boldmath $k$})\right]={Z_1\over L},
\end{equation}
where $L$ is now the number of lattice sites, and
$\varepsilon(\mbox{\boldmath $k$})$ is the non-interacting dispersion derived
from $K$. [In the limit of vanishing activity,
$\Psi(1)=L\ln(1+e^{\beta\mu}Z_1/L)\to e^{\beta\mu}Z_1$, which is the
classical non-interacting result.] The occupation probabilities (\ref{pi})
are then also independent of position, and to first order in the activity,
they become equal to the (dimensionless) fugacity:
\begin{equation}
p={e^{\beta\mu}Z_1/L\over 1+e^{\beta\mu}Z_1/L}\to
e^{\beta\mu}{Z_1\over L}\to e^{\beta\mu}\left(a\over\lambda_T
\right)^3,
\label{fug}
\end{equation}
where the second limit is of high temperature, with $a$ the lattice constant
and $\lambda_T$ the thermal wavelength.

The classical limit for $\Psi(2)$ is taken in the usual two steps: first, the
commutator $[K,V]$ is neglected, being at least of order $\hbar$, so that
disentanglement is trivial:
\begin{equation}
e^{-\beta(K+V)}\rightarrow e^{-\beta K}e^{-\beta V},
\end{equation}
after which the two-particle matrix element reads
\begin{eqnarray}
 U_{ij}&=&\left<\mbox{\boldmath $r$}_i,\mbox{\boldmath $r$}_j\left|
e^{-\beta K}e^{-\beta V}\right|\mbox{\boldmath $r$}_i,
\mbox{\boldmath $r$}_j\right>\nonumber\\
&&={1\over L^2}e^{-\beta v_{ij}}\left[Z_1^2-\left(
\sum_{\mbox{\boldmath $k$}}\cos\left[\mbox{\boldmath $k$}\cdot
(\mbox{\boldmath $r$}_i-\mbox{\boldmath $r$}_j)\right]
\exp\left[-\beta\varepsilon(\mbox{\boldmath $k$})\right]
\right)^2\right].
\end{eqnarray}
Here $v_{ij}=V(|\mbox{\boldmath $r$}_i-\mbox{\boldmath $r$}_j|)$, and the
interference term from $\left|\left< \mbox{\boldmath $r$}_i,\mbox{\boldmath
$r$}_j | \mbox{\boldmath $k$}_i,\mbox{\boldmath $k$}_j\right>\right|^2$ is
shown explicitly. In physical units $\mbox{\boldmath $k$}=\mbox{\boldmath
$p$}/\hbar$, so one observes in the second step that as $\hbar\to 0$, it gets
`washed out' by the sum, giving
\begin{equation}
\Psi(2)=\sum_{1\leq i<j\leq L}\ln\left[1+p^2\left(e^{-\beta v_{ij}}-1
\right)\right],
\end{equation}
where $p$ is the position-independent occupation probability (\ref{fug}),
multiplying Mayer's expansion parameter, $g_{ij}=e^{-\beta v_{ij}}-1$. In
other words, the subtracted matrix elements $S_{ij}$ in the probability
expansion become equal to Mayer's parameter: $S_{ij}\to g_{ij}$ in the
classical limit.

This is as far as one can go without invoking the probability expansion
(\ref{prob}) explicitly, since only $\Psi(2)$ is easy to rewrite in
probabilities `by hand'. It is not difficult to show that, when $\hbar\to 0$
and all interference terms are neglected, the normalized matrix elements
(\ref{normu}) in the position basis take the familiar form
\begin{equation}
\widetilde{U}_{j(n)}=\exp\left(-\beta\sum_{1\leq k<l\leq n}v_{j_kj_l}
\right)=\prod_{1\leq k<l\leq n}(1+g_{j_kj_l}),
\label{uw}
\end{equation}
where the sum (product) is over all pairs of indices in
$j(n)=\{j_1,\ldots,j_n\}$. It follows that $\Psi$ is a function of the
$g_{ij}$ alone, as is well known. The $S$'s vanish with the interaction, as
they should, because all the $\widetilde{U}$'s in equation (\ref{uw}) are
then equal to unity, so they cancel in the definition (\ref{sdef}).

This is a good place to pause, and put the result (\ref{uw}) in a physical
perspective. One could have worried: the same one-particle terms (\ref{onep})
appear in the denominator of the normalized matrix elements (\ref{normu}) as
in the numerator of the probabilities (\ref{pi}). Might they not cancel,
leaving the probabilities in expressions like (\ref{psi3}) only formally, but
not really, in control of the convergence? Not so: equation (\ref{uw}) shows
that one should rather expect one-particle (`kinetic') terms to cancel between
the numerator and denominator of $\widetilde{U}$ itself, this cancellation
being complete in the classical limit. This is just the statement, that in the
classical limit one can integrate out momenta from the partition function
exactly. Quantum effects are manifested as incomplete cancellation:
interactions do affect the momentum distribution. However, to believe that a
$\widetilde{U}$ would not tend to a limit when a one-particle matrix element
in the denominator became small, is to believe that the many-body state in the
numerator has an `infinite stopping power' for that particle, if it can avoid
making the numerator small as well, despite containing that same
single-particle state of high momentum. Such drastic effects of the
interaction on the momentum distribution are not unimaginable. In fact, one of
the persistent worries in high-temperature superconductivity is that they
could preclude any `semiclassical' description of the conducting
electrons~\cite{Anderson90}. Nevertheless, it may be (vaguely) concluded, that
only `exotic' collective states would spoil the \emph{numerical} convergence
of Kubo's expansion for the grand potential, as long as the occupation
probabilities (\ref{pi}) are reasonably small.

\subsection{The virial series}

Going back from probability to fugacity,
\begin{equation}
p={f\over 1+f},
\end{equation}
and expanding to second order in $f$, in the classical limit one obtains
\begin{equation}
\Psi(1)+\Psi(2)=Lf+\left(-{L\over 2}+\sum_{1\leq i<j\leq L}g_{ij}\right)f^2+
\mathcal{O}(f^3),
\end{equation}
and since it is consistent to write $g_{ii}=-1$, the term multiplying $f^2$
may be written as the \emph{unrestricted} double sum
\begin{equation}
{1\over 2}\sum_{i=1}^{L}\sum_{j=1}^{L}g_{ij}\to {L\over 2}
\sum_{\Delta} g(\Delta)\equiv LB_1,
\end{equation}
where it has been used that $g_{ij}\equiv g(|\mbox{\boldmath
$r$}_i-\mbox{\boldmath $r$}_j|)$ depends only on the differences, and the
arrow means the large-volume limit. It is obvious that $B_1$ is just the
usual `second virial coefficient', except that it is dimensionless; this can
be repaired by transferring the factor $a^3$ from the fugacity (\ref{fug}),
and putting $La^3=V$.

The third-order result is recovered along the same lines. All that remains of
$\Psi(3)$ (\ref{psi3}) to third order in the fugacity is $f^3 \sum S_{ijk}$,
where, in the classical limit [inserting (\ref{uw}) into (\ref{sdef})],
\begin{equation}
S_{ijk}=g_{ij}g_{ik}g_{jk}+g_{ij}g_{ik}+g_{ij}g_{jk}+g_{ik}g_{jk},
\label{S3}
\end{equation}
from which it is clear that the subtracted matrix elements $S_{i(n)}$ are the
generating functions of labelled graphs with $n$ vertices. Adding the terms in
$f^3$ from $\Psi(1)$ and $\Psi(2)$, one gets
\begin{equation}
\left({L\over 3}-2\sum_{i<j}g_{ij}+\sum_{i<j<k} S_{ijk}\right)f^3.
\label{vir3}
\end{equation}
To pass from restricted to unrestricted summation, use
\begin{equation}
3!\sum_{i<j<k}=\sum_{i\neq j\neq k}=\sum_{i,j,k}-\sum_{i,j=k}-\sum_{i=k,j}-\sum_{i=j,k}+
2\sum_{i=j=k},
\label{inv}
\end{equation}
which may be derived by inverting successive expressions of the type
\begin{equation}
\sum_{i,j}=\sum_{i\neq j}+\sum_{i=j}.
\end{equation}
Inserting (\ref{S3}) and (\ref{inv}) into (\ref{vir3}), it becomes
\begin{equation}
{1\over 6}\sum_{i,j,k}S_{ijk}f^3,
\end{equation}
so that, just as in second order, all that the lower terms in (\ref{vir3}) do
is to remove restrictions on the sum in the highest one. In the large-volume
limit, the first term in (\ref{S3}) gives
\begin{equation}
{1\over 6}\sum_{i,j,k}g_{ij}g_{ik}g_{jk}\to
{L\over 6}\sum_{\Delta_1,\Delta_2}g(\Delta_1)g(\Delta_2)
g(|\Delta_1-\Delta_2|)\equiv
LB_2,
\end{equation}
and the remaining three give a total contribution
\begin{equation}
{1\over 2}\sum_{i,j,k}g_{ij}g_{ik}\to{L\over 2}\sum_{\Delta_1}\sum_{\Delta_2}
g(\Delta_1)g(\Delta_2)=2 L B_1^2,
\end{equation}
because the $g$'s depend only on the differences, so one finds, finally,
\begin{equation}
\Psi=Lf+LB_1f^2+L(2B_1^2+B_2)f^3+\mathcal{O}(f^4),
\end{equation}
which is the virial series to third order, in one notation~\cite{Balian91}.

The statement that the interacting problem is `reduced to quadrature' in the
classical limit is interpreted here, that the coefficient of $f^n$ in $\Psi$
becomes an \emph{unrestricted} sum, namely
\begin{equation}
{1\over n!}\sum_{i_1,\ldots,i_n}
\left\{\mbox{coefficient of $f^n$ in $\psi[i(n)]$}\right\}.
\end{equation}
From this point of view, in the classical limit there appears a `conspiracy
of terms' which removes quantum restrictions from the sums in Kubo's formula.

\section{Discussion}

This article gives a combinatorial interpretation of Kubo's cluster cumulant
expansion, as a binomial transform of the grand potential. It primarily
explores the pedagogical implications of having such a short, but formal,
derivation. The idea is to develop a self-contained, general point of
departure to treat problems which require a formulation in real space,
assuming only undergraduate prior knowledge. In particular, it is found there
is no need to introduce cumulants explicitly at this level, in order to
produce operative size-consistent expressions. This was demonstrated in
detail for the classical limit.

In this approach, the distinction is kept between rearranging the series for
$\Psi$, which (effectively) resums different infinite classes of terms, and
generating the terms themselves, which requires the evaluation of matrix
elements. It becomes clear in principle, how different choices of expansion
parameter (probability, fugacity, coupling constant,\ldots) necessarily yield
different rules for which terms appear, and how either quantum entanglement,
or topological restrictions from the Hamiltonian, both of which change the
form of (\ref{uw}), can spoil the `conspiracy' by which the classical
expressions simplify.

The basic operational problem in quantum mechanics is to replace sums by
integrals, or, in more sophisticated language, to pass from functions defined
on sets (of quantum states) to functions of real numbers (parameters of the
Hamiltonian). This is trivial with unrestricted sums, which is the
combinatorial content of the simplification in the classical limit. The
converse is formally the most difficult problem of strong correlations: when
dynamical effects restrict a (multiple) sum to an `arbitrary' subset of
discrete states, \emph{i.e.} such that no ordering can be defined on it,
there is no controlled way to express the sum as an integral in the large
volume limit. Otherwise, the standard way to obtain integrals is to introduce
ordering by the time variable, leading to Feynman diagrams. The relationship
between the cluster cumulant expansion and the diagrammatic approach has been
discussed by Dunn~\cite{Dunn75}, for the case of a particular self-energy. He
showed that cutting the expansion off at $m$-th order amounted to calculating
all diagrams with at most $m$-fold momentum integrals exactly, and all others
approximately [the logarithms in $\Psi(m)$ necessarily generate diagrams to
infinite order].

On the other hand, rewriting Kubo's formula as a probability expansion,
equation (\ref{prob}), shows that convergence can be expected even when one
does not have the complete solution of the problem. All that is required is
that the probabilities in the chosen basis are bounded away from unity (and
zero). The prototype for this is precisely the real-space basis, because
position states are never stationary, due to the uncertainty principle.

Two other properties of the cluster expansion are readily obtained. First, it
was stressed by Kubo~\cite{Kubo62} that all described operations remain
exactly correct even if the `matrix elements' are not $c$-numbers, and the
exponential functions are replaced by various rules. This follows directly
from the fact that the binomial inversion is a formal identity. For example,
when the probability expansion can be written in Fock-space operators, the
Pauli principle is exactly preserved, even if one stops at first
order~\cite{Sunko96}. Second, if matrix elements are used, the Pauli
principle cannot be satisfied for the whole assembly of particles, as soon as
the expansion is cut off. Its form then indicates that antisymmetrization is
taken into account by an `inclusion-exclusion' procedure; for instance, the
same two-particle terms, involving $S_{ij}$, appear with opposite sign in
$\Psi(2)$ and $\Psi(3)$ [equation (\ref{psi3})]. So the cluster cumulant
series is expected to alternate, whenever many-body correlations are
important. This can be striking in practice~\cite{He91}.

To conclude, the second and fourth sections of this article give a compact
and hopefully useful introduction to an established general treatment of
correlations in real space. In the third section, it is shown that the
parameters controlling convergence can be interpreted as probabilities, and
their associated subtracted matrix elements appear as the basic building
blocks of more elaborate calculations.

\section{Acknowledgements}

Conversations with S. Bari\v si\'c and one with B. Gumhalter are gratefully
acknowledged. This work was supported by the Croatian Government under Project
0119256. 

\appendix

\section{The expansion for bosons}

The probability expansion (\ref{prob}) is formally the same for fermions and
bosons, only the definitions of the various quantities change. For bosons,
the occupation probabilities are
\begin{equation}
p_i=\frac{\sum_{k=1}^\infty e^{k\beta\mu}\left<i^k\right|e^{-\beta H}
\left|i^k\right>}
{\sum_{k=0}^\infty e^{k\beta\mu}\left<i^k\right|e^{-\beta H}
\left|i^k\right>},
\end{equation}
where $i^k$ means, $i$-th state, occupied by $k$ bosons. In the
non-interacting case, this reduces to the familiar
$e^{\beta(\mu-\varepsilon_i)}$, justifying the use of the term `occupation
probability'.

The normalized matrix elements (\ref{normu}) become, for bosons,
\begin{equation}
\widetilde{U}_{j(n)}=\frac{
\sum_{k_1,\dots,k_n=1}^\infty e^{(k_1+\ldots+k_n)\beta\mu}
\left<j_1^{k_1}\ldots j_n^{k_n}\right|e^{-\beta H}
\left|j_1^{k_1}\ldots j_n^{k_n}\right>
}{\left(\sum_{k_1=1}^\infty e^{k_1\beta\mu}\left<j_1^{k_1}\right|e^{-\beta H}
\left|j_1^{k_1}\right>\right)\cdots
\left(\sum_{k_n=1}^\infty e^{k_n\beta\mu}\left<j_n^{k_n}\right|e^{-\beta H}
\left|j_n^{k_n}\right>\right)
}.
\end{equation}
To lowest order in the activity, these expressions are of course equal to the
fermion ones, which accounts for the classical limit. Note, finally, that the
denominator of $\widetilde{U}_{j(n)}$ contains the `occupation numbers'
\begin{equation}
n_i=\sum_{k=1}^\infty e^{k\beta\mu}\left<i^k\right|e^{-\beta H}
\left|i^k\right>,
\end{equation}
in terms of which $p_i=n_i/(n_i+1)$, as with non-interacting bosons.

\end{document}